\documentclass[10pt]{article}

\usepackage[utf8]{inputenc}
\usepackage[T1]{fontenc}
\usepackage[english]{babel}
\usepackage{amsmath,amsthm,setspace, amsfonts, amssymb,graphics,multirow,enumerate}
\usepackage{todonotes}
\usepackage{graphicx,color}

\usepackage{algorithm}
\usepackage{algcompatible}

 \usepackage{pdfpages}
\usepackage{float}
\usepackage{listings}
 \usepackage{bbm}
\usepackage{calc}
\usepackage{array}
 \usepackage{authblk}
 \usepackage{natbib}

\begin{document}


\def\logc{L}
\newcommand{\rset}{\mathbb{R}}
\newcommand{\obs}{\mathsf{Y}}
\newcommand{\hid}{X}
\newcommand{\Xset}{\mathsf{X}}
\newcommand{\Xsigma}{\mathcal{X}}
\newcommand{\pscal}[2]{\left<#1,#2\right>}
\newcommand{\rmd}{\mathrm{d}}
\newcommand{\eqdef}{:=}
\newcommand{\EM}{\mathrm{EM}}
\newcommand{\barS}{\overline{S}}
\newcommand{\hatS}{\widehat{S}}
\newcommand{\Q}{\mathcal{Q}}
\newcommand{\Qt}{\widetilde{\mathcal{Q}}}
\newcommand{\calL}{\mathcal{L}}
\newcommand{\A}{\mathsf{A}}
\newcommand{\F}{\mathcal{F}}
\newcommand{\PE}{\mathbb{E}}
\newcommand{\1}{\mathbbm{1}}
\newcommand{\Wnorm}[1]{\left| #1 \right|_W}
\newcommand{\sqrtWnorm}[1]{\left| #1 \right|_{\sqrt{W}}}
\newcommand{\WnormM}[1]{\left\| #1 \right\|_W}
\newcommand{\sqrtWnormM}[1]{\left\| #1 \right\|_{\sqrt{W}}}
\newcommand{\D}{\mathsf{D}}

\newcommand{\coint}[1]{\left[#1\right)}
\newcommand{\ocint}[1]{\left(#1\right]}
\newcommand{\ooint}[1]{\left(#1\right)}
\newcommand{\ccint}[1]{\left[#1\right]}
\newcommand{\jacob}[2]{\operatorname{J} #1(#2)}
\def\argmin{\operatorname{argmin}}
\def\argmax{\operatorname{argmax}}
\def\Prox{\operatorname{Prox}}

\def\eqsp{\,}


\title{Fast selection of nonlinear mixed effect models using penalized likelihood}

\author[1]{Edouard Ollier}
\affil[1]{INSERM, U1059, Dysfonction Vasculaire et H\'emostase, Saint Etienne, France.}

\maketitle

\begin{abstract}
Nonlinear Mixed effects models are hidden variables models that are widely used in many fields such as pharmacometrics. In such models, the distribution characteristics of hidden variables can be specified by including several parameters such as covariates or correlations which must be selected. Recent development of pharmacogenomics has brought averaged/high dimensional problems to the field of nonlinear mixed effects modeling for which standard covariates selection techniques like stepwise methods are not well suited. The selection of covariates and correlation parameters using a penalized likelihood approach is proposed. The penalized likelihood problem is solved using a stochastic proximal gradient algorithm to avoid inner-outer iterations. Speed of convergence of the proximal gradient algorithm is improved using component-wise adaptive gradient step sizes. The practical implementation and tuning of the proximal gradient algorithm are explored using simulations. Calibration of regularization parameters is performed by minimizing the Bayesian Information Criterion using particle swarm optimization, a zero-order optimization procedure. The use of warm restart and parallelization allowed computing time to be reduced significantly . The performance of the proposed method compared to the traditional grid search strategy is explored using simulated data. Finally, an application to real data from two pharmacokinetics studies is provided, one studying an antifibrinolytic and the other studying an antibiotic.
\end{abstract}


\section{Introduction}

Nonlinear Mixed effects modelling \citep{lavielle2014mixed} is a classical approach to analyse longitudinal data. It has been extensively used in many fields such as pharmacometrics \citep{pillai2005non}, oncology \citep{desmee2016using, ollier2016analysis} and forestry \citep{xu2014nonlinear}. Nonlinear mixed effects models (NLMEM) belong to the class of latent variables models. The non-linearity of the regression function with respect to hidden variables makes the parameters estimation challenging, the likelihood being intractable. State-of-the-art methods use Expectation-Maximization (EM) type algorithms to perform parameters estimation. As the non-linearity of the regression function does not allow the E-step to be done in closed form, it is performed by using a Monte Carlo approximation which corresponds to the MCEM algorithm \citep{wei1990monte}. As MCEM algorithm requires the use of an important number of hidden variable simulation for each iteration of the EM algorithm, it is particularly computationally demanding. By introducing a Robbins-Monro approximation scheme, the Stochastic Approximation Expectation Algorithm (SAEM) \citep{kuhn2004coupling} allows the use of few Monte Carlo simulations per iteration and reduces drastically the computational cost.
\\$ $\\
Before performing the estimation of a NLMEM, the distributional characteristic of the latent variables must be specified. Correlation between latent variables and the influence of external source of information, the covariates, can be introduced. In this later case, the mean of each latent variable is then defined as a linear model which design matrix summarize all the external information. The choice of covariates/correlation that must be included in the model is of crucial interest.  For example, genotypic covariates can be introduced in tumour growth inhibition model to improve model predictiveness \citep{mazzocco2015prediction}. In population pharmacokinetic, the inclusion of a concomitant treatment as a covariate can be used to test the existence of a drug interaction \citep{panhard2005evaluation, delavenne2013semi}. The development of a NLMEM consists then in the selection of statistically influent correlations and/or covariates for each latent variable. Even if the great majority of problems are not high dimensional ones, the number of all the possible models might be very important. As an example, for a NLMEM with $L=4$ latent variables and $K=10$ potential covariates to study, the number of all possible the model is equal to $2^{L*K + \frac{L\times \left(L-1\right)}{2}}>7\times 10^{13}$ . Such dimension precludes the use of a greedy approach and makes automated selection methods appealing to reduce the dimension of the search space.  Moreover, the recent development of pharmacogenomic bring high dimensional problems to the field of NLMEM for which standard covariates selection techniques such as stepwise methods are not well suited.
\\$ $\\
As in classical statistical model like generalized linear model, the use of penalized estimator using sparsity inducing norm is an attractive solution. In the case of NLMEM, it is computationally difficult as it requires to solve a non-smooth intractable optimization problem. Some pioneer works already proposed extensions of the SAEM algorithm to solve such maximum likelihood problems penalized by different types of sparsity inducing penalties.  Bertrand et al. \citep{bertrand2013multiple} proposed the use of the $l_1$ norm to select covariates in pharmacogenomics studies but without calibrating the penalty parameters. Ollier et al. \citep{ollier2016saem} used the generalized fused lasso penalty to select between group differences of fixed effects and variance of random effect. Although the proposed algorithms showed good numerical properties, no theoretical work studied the statistical convergence of such extensions. Moreover, these $SAEM$ algorithm solves numerically a non-smooth optimization problem at each iteration. This inner/outer iteration scheme have a non-negligible computational cost. Recently, Atchade et al. \citep{atchade2014stochastic} and Fort et al. \citep{fort2017stochastic} extended proximal gradient methods \citep{beck2009gradient} to problems with intractable likelihood. It is based on a proximal gradient scheme with a gradient numerically approximated using a Monte Carlo method, avoiding the use of an inner/outer iteration.
\\$ $\\
Different additional problems complicated the development of penalized methods for NLME. Selection of correlation parameters by directly penalizing the elements of the covariance matrix faces two main issues. First, as covariances between random effects depends on their respective variances, scaling issues may appear when trying to select the covariance matrix of random effects with large differences in variance values. Secondly, performing directly a gradient based optimization of the covariance matrix is a complicated task as the new estimates at each iteration must be positive semi-definite. To cope with these issues, it has been proposed to use a modified Cholesky decomposition \citep{chen2003random, bondell2010joint} of the covariance matrix. This decomposition allows the separate estimation of parameters related to variances and correlation of random effects.
\\$ $\\
Like with other penalized likelihood approach, the optimal value of penalty parameters must be estimated using a criterion like cross-validation error \citep{colby2013cross} or Bayesian Information Criterion (BIC) \citep{delattre2014note}. This step relies generally on the evaluation of a user-defined grid of penalty parameters values, the optimal value being the one that minimizes the criterion used for selection. Such an approach faces two issues: i) it is time-consuming for a grid with a numerous elements and ii) it makes the results of the selection process highly dependent on the values of penalty parameters provided by the user. To make this calibration faster and less user dependent, a routine that automatically estimates the optimal value of the penalty parameters is needed. Furthermore, this routine should not rely on gradient based methods as the gradient of selection criteria with respect to the penalty parameters are intractable. Among zero order optimization methods that may be used, particle swarm optimization algorithm \citep{kennedy1995particle}, is an interesting solution as it relies on simple formula without any inner optimization step unlike Bayesian optimization \citep{jones1998efficient} and can be easily parallelized. Particle swarm optimization has already been used in various applications such as chemometrics \citep{shen2004modified} or biomedical signal analysis \citep{ince2009generic}.
\\$ $\\
A stochastic proximal gradient algorithm for the selection of both covariates and correlation parameters is presented. The selection of the optimal value for the regularization parameters is automatically performed by minimizing the BIC using a zero-order optimization method, particle swarm optimization. Section \ref{NLME} introduces NLMEM, the penalized maximum likelihood problem used to jointly select correlation and covariates parameters is introduced in section  \ref{PenLik}. In section \ref{SAPG}, the stochastic proximal gradient algorithm specifically developed to solve such penalized likelihood problems is described. Calibration of penalty parameters is discussed in section \ref{PenParam}. In section \ref{sec:sim}, the proposed algorithm is evaluated on simulated data. Finally, a real data application focusing on the population pharmacokinetics of an antifibrinolitic drug, and an antibiotic is provided in section \ref{sec:realdata}.

\section{Non-linear mixed effects models} \label{NLME}

 \subsection{Generalities on Non-linear mixed effects models}
Let us assume that we observe longitudinal data along time for $N$ patients.
 Let $Y_{i,j}$ be the observation at time $t_{i,j}$ ($j \in \{1,\ldots,J \}$)
 for the $i$-th patient ($i \in \{1,\ldots,N\}$). The mixed model is written as
 \begin{eqnarray}
&& Y_{i,j} = f(t_{i,j},Z_{i}) +  \epsilon_{i,j}  \nonumber ,\\
&& \epsilon_{i,j} \sim \mathcal{N}(0,\sigma^2) \mbox{ (iid)}, \nonumber
\end{eqnarray}
where $f$ is a given non-linear function.   Measurement errors $\epsilon_{i,j}$ are further assumed to be independent and identically distributed with variance $\sigma^2$. Individual parameters $Z_{i}$ for the $i$-th subject are $L$-dimensional random vectors, independent of $\epsilon_{i,j}$ and  assumed to be distributed  (up to a transformation $h$) as:
\begin{eqnarray}
 h(Z_{i}) &\sim& \mathcal{N}(X_{i} \beta, \Omega) \mbox{ (iid)}, \nonumber
\end{eqnarray}
where $\beta \in \mathbbm{R}^{(K+1)L}$ is the mean parameter  vector. Various transformations $h$ can be used. A common one $h(z) = \left( \log(z_{1})  , \dots , \log(z_{L})  \right) $, which yields log-normally distributed $Z_i$.  The matrix $X_i \in \mathbbm{R}^{L \times (K+1)L}$ corresponds to the design matrix of the $i$-th subject and is of the form: 
\begin{eqnarray}
X_{i} = \begin{pmatrix} ( 1, x_i^1 , \dots , x_i^K) & 0 & \dots & 0 \\  \vdots & &  & \vdots \\ 0 & \dots & 0 & ( 1 , x_i^1 , \dots , x_i^K)  \end{pmatrix} \nonumber.
\end{eqnarray}
Where $x_i^k$ corresponds to the value of the $k$-th ($k \in \{1,...,K \}$) covariate for the $i$-th patient. Finally, the vector of unknown parameters is $\theta = \left(\beta, (\Omega_{kk^{'}})_{1 \leq k\leq k^{'}\leq \textcolor{red}{L}}, \sigma \right)\in \mathbb{R}^D$.

\medskip
The mixed model is a latent variable model because the individual random variable $\{Z_i, \mbox{ } i=1,...,N \}$ are unobserved.
The log-likelihood  is:
\begin{gather}\label{eq:likelihood}
\ell(\theta) = \sum_{i=1}^{N} \log  p(Y_{i};\theta)= \sum_{i=1}^{N}   \log \left(   \int p(Y_{i} ,Z_i ; \theta) \rmd Z_i \right),
\end{gather}
where  $Y_{i} \in \mathbbm{R}^{J}$ and $p(Y_{i},Z_i ; \theta)$ are respectively the vector of observations and  the likelihood of the complete data of the $i$-th subject :
 \begin{eqnarray*}
\log p(Y_{i},Z_i ; \theta) &\propto&    - J \log(\sigma )-\frac{1}{2}\sum_{j}\frac{ \left( Y_{ij} - f(t_{i,j},Z_i)\right)^{2} }{\sigma^2}  - \frac{1}{2}\log(\vert \Omega \vert)  - \frac{1}{2} (h(Z_i) -  X_{i}  \beta)^{t}\Omega^{-1}(h(Z_i) -   X_{i}  \beta),   \nonumber \\
&\propto&    - J \log(\sigma )-\frac{1}{2}\sum_{j}\frac{ \left( Y_{ij} - f(t_{i,j},Z_i)\right)^{2} }{\sigma^2}  - \frac{1}{2}\log(\vert \Omega \vert)  - \frac{1}{2} \mbox{Trace} \left(  \Sigma\textcolor{red}{_i} \Omega^{-1}\right),  \nonumber
\end{eqnarray*}

with $\Sigma_i =   h(Z_i)h(Z_i)^t - h(Z_i) (X_{i}  \beta)^t - (X_{i}  \beta) h(Z_i)^t + (X_{i}  \beta) (X_{i}  \beta)^t  $.

Note that the complete likelihood belongs to the  curved exponential family and therefore can be written as:
$$\log p(Y_{i},Z_i ; \theta) =   \varphi(\theta) +    \pscal{S(Z_i)}{\psi(\theta)}, $$
 where  the minimal sufficient statistics $S(Z_i)$  
are
$S_{1}(Z_i) =  h(Z_i)$, $S_{2}(Z_i) =  h(Z_i) \, h(Z_i)^t$ and
$S_{3,i}(Z_i) = \sum_{j=1}^{J} (Y_{ij}-f(t_{ij}, Z_i))^2$. Using these sufficient statistics, the complete log-likelihood takes the form:
 \begin{eqnarray} \label{eq:compll}
\log p(Y_i,S(Z_i) ; \theta) \propto    - J \log(\sigma )-\frac{1}{2}\frac{ S_{3,i}(Z_i) }{\sigma^2}  - \frac{1}{2}\log(\vert \Omega \vert)  - \frac{1}{2} \mbox{Trace} \left(  \Sigma_i \Omega^{-1}\right), 
\end{eqnarray}
with $\Sigma_i =     S_{2}(Z_i) - S_{1}(Z_i) (X_{i}  \beta)^t - (X_{i}  \beta) S_{1}(Z_i)^t + (X_{i}  \beta) (X_{i}  \beta)^t  $. The complete log-likelihood can be calculated up to an additive constant using the sufficient statistics $S_1(Z_i)$, $S_2(Z_i)$, $S_{3,i}(Z_i)$  and the vector $\theta$.

 \subsection{Reparametrization of covariance matrix}
 Performing directly a gradient based optimization of the covariance matrix $\Omega$ is a complicated task as the new estimates at each iteration of the algorithm must be positive semi-definite. To cope to these issues, it has been proposed to use a modified Cholesky decomposition \citep{chen2003random, bondell2010joint} of the covariance matrix. This decomposition allows the separate estimation of parameters related to variances and correlations of random effects. Matrix $\Omega$ can be decomposed using a modified decomposition \citep{chen2003random}:  
  \begin{eqnarray}
  \Omega = \Delta \Gamma \Gamma^t \Delta,
   \end{eqnarray}
   where $\Delta$ is a diagonal matrix with positive entries and $\Gamma$ a lower triangular matrix with 1's on the diagonal. The likelihood of the complete data (\ref{eq:compll}) can then be parametrized in the following way :
  \begin{eqnarray}
\log p(Y_i,S(Z_i) ; \theta) \propto    - J N \log(\sigma )-\frac{1}{2}\frac{ S_{3,i}(Z_i) }{\sigma^2}  - N\log(\vert \Delta \vert)  - \frac{1}{2} \mbox{Trace} \left(  \Sigma_i \left( \Delta \Gamma \Gamma^t \Delta \right)^{-1}\right). 
\end{eqnarray}
The vector of unknown parameters $\theta$ corresponds now to $\theta = \left(\beta,  (\Delta_{kk})_{1 \leq k \leq \textcolor{red}{L}},  (\Gamma_{k k^{'}})_{1 < k < k^{'} < \textcolor{red}{L}}, \sigma \right)\in \mathbb{R}^D$.

 \section{Penalized likelihood problem for non-linear mixed effects models selection} \label{PenLik}
When estimating a non-linear mixed effects model, it is often of interest to select influential parameters. The goal of the selection procedure is then to set to zero components of $\theta$ that are non-influential. This can be done by solving a penalized maximum likelihood problem of the form: 
 \begin{gather}\label{eq:PenLLgeneral}
\hat{\theta} =  \underset{\theta \in \mathbb{R}^d } {\operatorname{argmax}} \mbox{ }   \ell(\theta) - g(\theta),
\end{gather}
where the function $g(\theta)$ correspond to the penalty function and has the property to induce sparsity in the estimation of vector $\theta$. The type of penalty function $g(\theta)$ will depend on the quantity to select \citep{bach2011convex}. As the goal is to select covariates and non-diagonal elements of $\Omega$, we propose to solve the following penalized maximum likelihood problem:
\begin{equation}\label{eq:PenLL}
\begin{split}
& \hat{\theta}_{\lambda} =  \underset{\theta \in \mathbb{R}^d }{\operatorname{argmax}} \mbox{ }   \ell(\theta)  -    \lambda_\beta \Vert \beta \Vert_1 - \lambda_\Gamma \Vert \Gamma_{-} \Vert_{1} \\
& \mbox{ subject to: } \sigma > 0, \quad \Delta \succ 0 
\end{split},
\end{equation}
where $\Gamma_{-}$ is the lower triangular part of matrix $\Gamma$ ($\Vert \Gamma_{-} \Vert_{1} = \sum_{1\leq k < k^{'} \leq L} \vert \Gamma_{k k^{'}} \vert$). The constraint $\Delta \succ 0$ means that $\Delta$ needs to be a positive definite matrix. The regularization parameters $\lambda_\beta$ and $\lambda_\Gamma$ tune the penalty strength and so the level of sparsity obtained in the estimations of  $\beta$ and $\Gamma$.

  \section{Solving the penalized maximum likelihood problem} \label{SAPG}

\subsection{The Stochastic Approximation Proximal Gradient Algorithm}
 The penalized maximum likelihood problem (\ref{eq:PenLLAdapt}) can be solved using the Stochastic Approximation Proximal Gradient Algorithm (SAPG) \citep{fort2017stochastic}. The SAPG algorithm belongs to the family of proximal gradient algorithms \citep{combettes:pesquet:2011}. They are designed to solve optimization problems like problem (\ref{eq:PenLLgeneral}) and corresponds to the following iterative scheme:
 \begin{equation}
  \label{eq:PG}
 \theta_{n+1} =   \Prox_{g,\gamma_{n+1}}(\theta_n + \gamma_{n+1} \nabla_\theta \ell(\theta_n ) ),
  \end{equation}
where $\Prox_{g}(\theta)$ corresponds to the proximal operator associated to the penalty function $g$.

For penalties that take the form $g(\theta) = \lambda \Vert \theta \Vert_1$ with $\theta=(\theta_1, \cdots, \theta_d)$ and $\lambda >0$, this proximal operator is the soft-thresholding operator. For any component
$i \in \{1, \cdots, d \}$, it corresponds to:

\[
 \left( \Prox_{\gamma,g}(\theta) \right)_i =  \left\{
  \begin{array}{ll}
    \theta_i -  \gamma \lambda & \text{if $\theta_i \geq \gamma \lambda $,} \\
    0 & \text{if $ |\theta_i| \leq \gamma \lambda $,} \\
\theta_i + \gamma \lambda  & \text{if $\theta_i \leq - \gamma \lambda $.}
  \end{array}
\right.
\]
  
Unfortunately, when dealing with nonlinear models, the gradient $\nabla_\theta \ell(\theta)$ has no explicit form. Using the Fisher identity, the gradient of the log-likelihood $\ell(\theta)$ is given by :
$$\nabla_\theta \ell(\theta_n ) = \mathbbm{E}_{\theta_n } \left[ \sum_{i=1}^{N} \nabla_\theta \log p(Y_i,S(Z_i) ; \theta_n )   \bigg \vert Y_i  \right], $$
under the conditional distribution $p(Z \vert Y ; \theta_n)$. For a model belonging to the curved exponential family, this corresponds to compute the conditional expectation of the complete data sufficient statistics: 
$$\bar{S}_{1,i}(\theta_n) =  \mathbbm{E}_{\theta_n} \left [ h(Z_i) \right \vert Y_i], \quad \bar{S}_{2,i}(\theta_n) =  \mathbbm{E}_{\theta_n} \left[  h(Z_i)h(Z_i)^t \bigg \vert Y_i \right], \quad   \bar{S}_{3,i}(\theta_n) = \mathbbm{E}_{\theta_n} \left[  \sum_{j=1}^{J} (Y_{ij}-f(t_{ij}, Z_i))^2 \bigg \vert Y_i \right].$$  
Sufficient statistics depend also on $Y_i$, but for convenience this notation is omitted. We then have:
\begin{eqnarray*}
 &&  \mathbbm{E}_{\theta_n} \left[ \sum_{i=1}^{N} \frac{\partial \log p(Y_i,S(Z_i) ; \theta_n ) }{\partial \beta}  \bigg \vert Y_i   \right] =    \sum_i  X_{i}^t \Omega^{-1}_n \left( \bar{S}_{1,i}(\theta_n) - X_{i}  \beta_n \right), \nonumber \\
 &&  \mathbbm{E}_{\theta_n} \left[ \sum_{i=1}^{N} \frac{\partial \log p(Y_i,S(Z_i) ; \theta_n ) }{\partial \Gamma}  \bigg \vert Y_i   \right] = \left( \Gamma \Gamma^t \right)^{-1} \left( \Delta^{-1} \bar{\Sigma}_n \Delta^{-1} \right)  \left( \Gamma \Gamma^t \right)^{-1} \Gamma, \nonumber \\
  &&  \mathbbm{E}_{\theta_n} \left[ \sum_{i=1}^{N} \frac{\partial \log p(Y_i,S(Z_i) ; \theta_n ) }{\partial \Delta}  \bigg \vert Y_i   \right] = N \Delta^{-1} + \Delta^{-1} \left( \Gamma \Gamma^t \right)^{-1}  \left( \Delta^{-1} \bar{\Sigma}_n \Delta^{-1} \right),  \nonumber \\
 &&  \mathbbm{E}_{\theta_n} \left[ \sum_{i=1}^{N} \frac{\partial \log p(Y_i,S(Z_i) ; \theta_n ) }{\partial \sigma}   \bigg \vert Y_i   \right] = -\frac{NJ}{\sigma_n} + \frac{ \sum_i \bar{S}_{3,i}(\theta_n)}{\sigma_n^3}, \nonumber
\end{eqnarray*}

and $\bar{\Sigma}_n =  \sum_{i=1}^{N}  \left( \bar{S}_{2,i}(\theta_n) - \bar{S}_{1,i}(\theta_n) (X_{i}  \beta_n)^t - (X_{i}  \beta_n) \bar{S}_{1,i}(\theta_n)^t + (X_{i}  \beta_n) (X_{i}  \beta_n)^t  \right)$. But, $\bar{S}_{1,i}(\theta_n)$, $\bar{S}_{2,i}(\theta_n)$ and $\bar{S}_{3}(\theta_n)$ have not an explicit form and then have to be numerically approximated. In the SAPG algorithm, this approximation is done using the following stochastic approximation scheme:  
\begin{eqnarray*}
&&  S^{sa}_{1,i, n+1} = S^{sa}_{1,i, n} + \delta_{n+1} \left(  \frac{1}{M_n} \sum_{m=1}^{M_n} S_{1,i}(Z_{m,n}) -  S^{sa}_{1,i, n}  \right) \approx \bar{S}_{1,i}(\theta_n)  \nonumber \\
&&  S^{sa}_{2,i, n+1} = S^{sa}_{2,i, n} + \delta_{n+1} \left(  \frac{1}{M_n} \sum_{m=1}^{M_n} S_{2,i}(Z_{m,n}) -  S^{sa}_{2,i, n}  \right) \approx \bar{S}_{2,i}(\theta_n) \nonumber \\
&&  S^{sa}_{3,i, n+1} = S^{sa}_{3,i, n} + \delta_{n+1} \left(  \frac{1}{M_n} \sum_{m=1}^{M_n} S_{3,i}(Z_{m,n}) -  S^{sa}_{3,i, n}  \right)\approx \bar{S}_{3,i}(\theta_n), \nonumber 
\end{eqnarray*}
with $Z_{1,n},...,Z_{M_n,n}$ a sequence of individual parameters simulated using a Monte Carlo Markov Chain (MCMC) algorithm with conditional distribution $p(Z \vert Y, \theta_n)$ \citep{kuhn2004coupling}. The sequence $\{ \delta_n, n \geq 0\}$ corresponds to the stochastic approximation step sizes. A classical choice \citep{fort2017stochastic} is:
\begin{equation}
  \label{eq:deltaSA}
  \delta_{n+1}   = \left\{ 
\begin{array}{ll} 
  \delta_\star  & \text{ if } n \leq n_{sa}, \\
  \delta_\star (n-n_{sa})^{-\zeta}  & \text{ if } n>n_{sa}.
\end{array} \right. 
\end{equation}
$\delta_n$ is then set to a constant value $\delta_\star$ during the first $n_{sa}$ iterations and then decrease with a rate tuned by parameters $\zeta \in \mathbbm{R}^{+} $. Hyperparameters $n_{sa}$ and $\zeta$ were set respectively to 0 and 0.75. As Monte Carlo updates correspond to a step with high computational cost, it is possible to perform MCMC updates not at each iteration but according to fixed frequency (Freq, see algorithm \ref{algo:SAPG}).

By using these approximations, we get $G^{sa}(\theta_n)$ an approximation of $\nabla_\theta \ell(\theta_n)$ with:
\begin{eqnarray*}
 && G^{sa}_\beta (\theta_n) = \sum_i  X_{i}^t \Omega^{-1}_n \left( S^{sa}_{1,i, n+1} - X_{i}  \beta_n \right) \nonumber \\
 && G^{sa}_\Gamma (\theta_n) =   \left( \Gamma_n \Gamma_n^t \right)^{-1} \left( \Delta_n^{-1} \Sigma_n^{sa} \Delta^{-1}_n \right)  \left( \Gamma_n \Gamma_n^t \right)^{-1} \Gamma_n  \nonumber \\
  && G^{sa}_\Delta (\theta_n) =  N \Delta^{-1}_n + \Delta^{-1}_n \left( \Gamma_n \Gamma^t_n \right)^{-1}  \left( \Delta_n^{-1} \Sigma_n^{sa} \Delta_n^{-1} \right) \nonumber \\
 && G^{sa}_\sigma (\theta_n) = -\frac{NJ}{\sigma_n} + \frac{\sum_i S^{sa}_{3,i, n+1}}{\sigma_n^3}. \nonumber
 \end{eqnarray*}
with $\Sigma^{sa}_n =  \sum_{i=1}^{N}  \left( S^{sa}_{2,i} - S^{sa}_{1,i} (X_{i}  \beta_n)^t - (X_{i}  \beta_n) S^{sa^t}_{1,i} + (X_{i}  \beta_n) (X_{i}  \beta_n)^t  \right)$. By replacing $\nabla_\theta \ell(\theta_n)$ by its stochastic approximation $G^{sa}(\theta_n)$ in equation (\ref{eq:PG})  we get the iterative scheme of the SAPG algorithm:
$$\theta_{n+1} =   \Prox_{g,\gamma_{n+1}}( \theta_n + \gamma_{n+1} G^{sa}(\theta_n) ).$$

\subsection{Gradient step size tuning strategies} \label{sec:algo}
As in every gradient-based optimization procedure, step size values influence greatly the speed of convergence. Moreover, in the case of NLME, parameters of different nature are optimized (mean, variances, correlations) and using the same step size sequence for all the parameters may be not suitable. This section presents two-step size tuning strategies.

\subsubsection{Common step size sequence}
A first and simple strategy is to use for all the parameters the same sequence $\{ \gamma_{n}; n \geq 0\}$ of the form:
\begin{equation}
  \label{eq:gammaCst}
  \gamma_{n}   = \left\{ 
\begin{array}{ll} 
  \gamma_0  & \text{ if }  n \leq n_\alpha, \\
  \gamma_0 (n-n_\alpha)^{-\alpha}  & \text{ if } n >n_\alpha.
\end{array} \right.
\end{equation}
Hyperparameters $n_\alpha$ and $\alpha$ were set respectively to 0 and 0.5. In practice using $n_\alpha>0$ may improve the convergence speed, but we found that choosing  $n_\alpha=0$ was a more robust solution when calibration of $\gamma_0$ was not accurate.

\subsubsection{Component-wise adaptive step size sequence}
In practice, the gradient vector may have very different scales depending on the coordinate of $\theta$. The step size $\gamma_n$ needs then to be adapted to these scales to ensure a fast convergence. In a previous work, Fort et al. \cite{fort2017stochastic} proposed to tune automatically the gradient step size using an approximation of the second derivatives of $\ell(\theta)$. This second order strategy provides a component-wise adaptive step size but necessitate extra computation that increase the computational cost of the algorithm. Moreover, it induces extra variability due to the approximation error of the second derivatives. Adaptive strategies based on first order information have been proposed in machine learning for models for which the computation of the second derivatives is expensive, like deep neural network. These strategies use a component wise adaptive step size sequence like the AdaGrad \citep{duchi2011adaptive} updating rule that is a simple strategy relying only on first order information. The sequence $\{ \gamma_{n}; n \geq 0\}$ is replaced by a vector-valued random sequence $\{\Lambda_{n} \in \mathbb{R}^D; n\geq 0\}$. The new update rule of SAPG is
\begin{gather*}
  \theta_{n}  = \Prox_{g,\Lambda_{n}} \left( \theta_{n} +  \Lambda_{n}   \circ G^{sa}(\theta_{n}) \right) = \left( \mbox{ }\Prox_{\lambda_d \Vert \Vert_1,\Lambda_{n}^{d}} \left( \mbox{ } (\theta_{n})_d +  \Lambda_{n}^{d}   \times (G^{sa}(\theta_{n}))_d \mbox{ } \right) \mbox{ } \right)_{d=1,...D}.
\end{gather*}
Where the vector $\Lambda^{n}$ is a vector with entries $\Lambda_{n}^{d}$:
\begin{equation}
\label{eq:gammaRandom}
 \Lambda_{n}^{d}  = \frac{\gamma_0}{\sqrt{ H_{n}^{d} + cst}},
  \end{equation}
with $cst$ a small positive constant that avoids division by zero (usually on the order of  $1e^{-8}$) and $H_{n}^{d}$ the sum of squared gradient:
\begin{equation}
  H_{n}^{d} = \sum_{k=1}^{n} (G^{sa}(\theta_{k}))_{d}^2.
  \end{equation}

\begin{algorithm}[H]
\caption{Stochastic Approximation Proximal Gradient algorithm}
\begin{algorithmic}[1]
\STATE \textbf{Initialization:} Set  $\beta_{1}$, $\Gamma_{1}$, $\Delta_{1}$, $\sigma_{1}$, $\delta_\star$, $\zeta$, $N_{SAPG}$, $Freq$
\FOR{$n \leq N_{SAPG}$}
\STATE 
\IF{(n=1) or (n mod Freq = 0)}
\STATE \emph{\textbf{Simulation step}} 
\STATE $\quad \quad$  Sample a path $Z_{1,n}, \cdots, Z_{M_{n},n}$ form the conditional distribution $p(Z \vert Y, \theta_n)$ and started from $Z_{M_{n},n-1}$ using a MCMC algorithm
\STATE \emph{\textbf{Expectation step}}  
\STATE Compute an approximation $S^{sa}_{n+1}$ of $\barS(\theta_n)$ 
\STATE $\quad \quad$  $\delta_{n} =  \delta_\star \left( \frac{n}{Freq}\right)^{-\zeta} $
\STATE $\quad \quad$  $S^{sa}_{n+1} = S^{sa}_{n} + \delta_{n} \left(  \frac{1}{M_n} \sum_{m=1}^{M_n} S(Z_{m,n}) -  S^{sa}_{n}  \right)$ 
\ELSE 
\STATE $\delta_{n} = \delta_{n-1}$
\STATE $S^{sa}_{n+1} = S^{sa}_{n}$  
\ENDIF
\STATE 
\STATE \emph{\textbf{Gradient computation}} 
\STATE $\quad \quad$  $G_\beta^{sa} (\theta_n) = \sum_i  X_{i}^t \Omega^{-1}_n \left( S^{sa}_{1,i, n+1} - X_{i}  \beta_n \right) $
\STATE $\quad \quad$  $G_\Gamma^{sa} (\theta_n) =   \left( \Gamma_n \Gamma_n^t \right)^{-1} \left( \Delta_n^{-1} \Sigma_n^{sa} \Delta^{-1}_n \right)  \left( \Gamma_n \Gamma_n^t \right)^{-1} \Gamma_n  $
\STATE $\quad \quad$  $G_\Delta^{sa} (\theta_n) =  N \Delta^{-1}_n + \Delta^{-1}_n \left( \Gamma_n \Gamma^t_n \right)^{-1}  \left( \Delta_n^{-1} \Sigma_n^{sa} \Delta_n^{-1} \right) $
\STATE $\quad \quad$  $G_\sigma^{sa} (\theta_n) = -\frac{NJ}{\sigma_n} + \frac{S^{sa}_{3, n+1}}{\sigma_n^3}$
\STATE 
\STATE \emph{\textbf{Step size computation}} 
\STATE $\quad \quad$  $\Lambda_{n}^{\beta} =  \frac{\gamma_0}{\sqrt{ H^{\beta}_{n} + cst }}$ with $H^{\beta}_{n} = H^{\beta}_{n-1} + G_{\beta}^{sa} (\theta_n)^2 $
\STATE $\quad \quad$  $\Lambda_{n}^{\Gamma} =  \frac{\gamma_0}{ \sqrt{ H^\Gamma_{n} + cst}}$ with $H^\Gamma_{n} = H^\Gamma_{n-1} + G_\Gamma^{sa} (\theta_n)^2 $
\STATE $\quad \quad$  $\Lambda_{n}^{\Delta} =  \frac{\gamma_0}{ \sqrt{ H^\Delta_{n} + cst}}$ with $H^\Delta_{n} = H^\Delta_{n-1} + G_\Delta^{sa} (\theta_n)^2 $
\STATE $\quad \quad$  $\Lambda_{n}^{\sigma} =  \frac{\gamma_0}{ \sqrt{ H^\sigma_{n} + cst}}$ with $H^\sigma_{n} = H^\sigma_{n-1} + G_\sigma^{sa} (\theta_n)^2 $
\STATE
\STATE \emph{\textbf{Proximal gradient step}} 
\STATE $\quad \quad$  $\beta_{{n+1}} = \Prox_{\lambda_\beta \Vert \cdot \Vert_1, \Lambda_{n}^{\beta}}\left( \beta_{n} + \Lambda_{n}^{\beta} \circ G_{\beta}^{sa} (\theta_{n}) \right)$ 
\STATE $\quad \quad$  $\Gamma_{n+1} = \Prox_{\lambda_\Gamma \Vert \cdot \Vert_1, \Lambda_{n}^{\Gamma}}\left( \Gamma_{n} + \Lambda_{n}^{\Gamma} \circ G_{\Gamma}^{sa} (\theta_{n}) \right)$ 
\STATE $\quad \quad$  $\Delta_{n+1} =  \Delta_{n} + \Lambda_{n}^{\Delta} \circ G_{\Delta}^{sa} (\theta_{n}) $ 
\STATE $\quad \quad$  $\sigma_{n+1} =  \sigma_{n} +\Lambda_{n}^{\sigma} \times G_{\sigma}^{sa} (\theta_{n}) $ 
\STATE 
\STATE \emph{\textbf{Projection step}} 
\STATE $\quad \quad$   $\Delta_{n+1} = \mbox{Proj}_{ \{ X \in \mathbbm{R}^{ L \times L } \vert X \succ  0  \} }(\Delta_{n+1})$ 
\STATE $\quad \quad$   $\sigma_{n+1} = \mbox{Proj}_{ \{ v \in \mathbbm{R}^{+} \vert v >  0  \} }(\sigma_{n+1})$ 
\STATE
\ENDFOR
\end{algorithmic}
\label{algo:SAPG}
\end{algorithm}

The SAPG  algorithm with component-wise adaptive step sizes  is summarized in Algorithm \ref{algo:SAPG}.

\begin{figure}[H]
\begin{center}
\includegraphics[scale = 0.5]{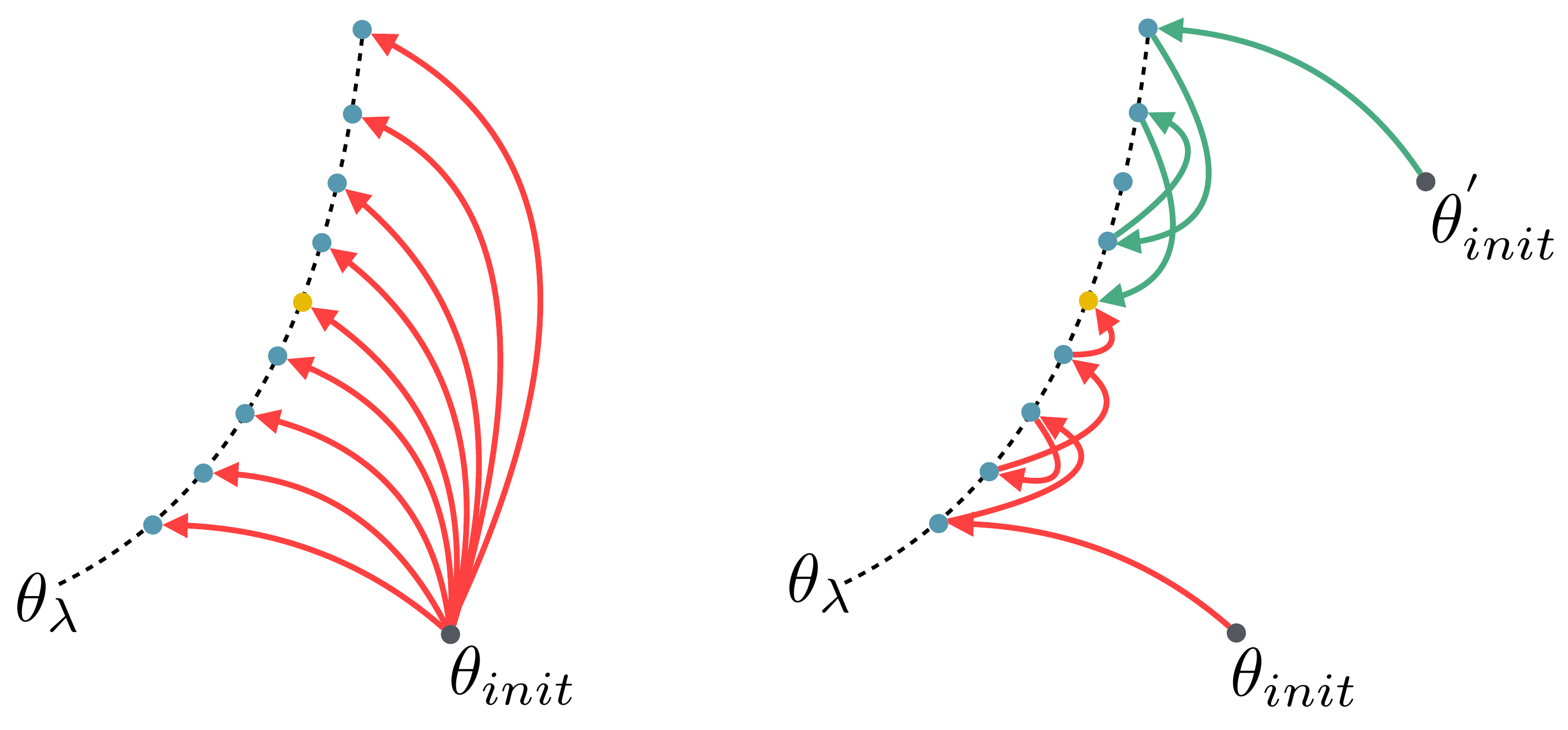}
\caption{Schematic representation of heuristics for the calibration of regularization parameter $\lambda$ : grid search (left) and particle swarm optimization (right). Blue dots represent the solutions of the penalized likelihood problem for different regularization parameter values. Yellow dots represent the solution of the penalized likelihood problem for the optimal regularization parameter values according to a selection criterion like BIC.}
\label{fig:SCHEMA_WR_PSO}
\end{center}
\end{figure}

 \section{Calibration of regularization parameter $\lambda$} \label{PenParam}
 
 \subsection{Selection criterion}

The calibration of regularization parameters $\lambda = (\lambda_\beta,\lambda_\Gamma)\in \mathbbm{R}^{2}$ is an important issue of penalized likelihood problems. A like selection criterion like the Bayesian Information criterion ($BIC$) \citep{delattre2014note} can be used:
\begin{eqnarray}
BIC(\hat{\theta}) = -2 \times \ell(\hat{\theta}) +  \log(N) \times \left( \vert Supp(\hat{\beta}) \vert + \vert Supp(\hat{\Gamma}) \vert  \right), \nonumber
\end{eqnarray}
where $\vert Supp(\hat{\beta}) \vert$ and $\vert Supp(\hat{\Gamma}) \vert$ correspond respectively to the size of the support of $\hat{\beta}$ and $\hat{\Gamma}$. 

As log-likelihood $\ell(\hat{\theta})$ has no closed form expression, it is approximated numerically using importance sampling \citep{samson2007saem}.  In practice, the SAPG algorithm could be run on a user-defined grid $(\lambda_{1},\ldots, \lambda_{M})$ to get a sequence of penalized estimates $\{ \hat{\theta}_{\lambda_{1}},\ldots, \hat{\theta}_{\lambda_{M}} \}$.  Then the optimal value  $\lambda_{BIC}$  is given by:
\begin{gather*}
\lambda_{BIC} =  \underset{ \lambda \in \{\lambda_{1},\ldots, \lambda_{M} \} } {\operatorname{ArgMin}} \mbox{ } BIC(\tilde{\theta}_{\lambda}), \nonumber
\end{gather*}
The BIC is computed using a two-step procedure in which the support of $\hat{\theta}_{\lambda}$ is re-estimated without penalty in order to remove the bias induced by the penalty in $\hat{\theta}_{\lambda}$. The vector $\tilde{\theta}_{\lambda}$ corresponds to this unpenalized re-estimation of selected (non-zero) coefficients in $\hat{\theta}_{\lambda}$. This grid based approach correspond to the left scheme of figure \ref{fig:SCHEMA_WR_PSO} as for each element of the user-defined grid $(\lambda_{1},\ldots, \lambda_{M})$ the SAPG algorithm is run from the same initial point $\theta_{init}$.

 \subsection{Automatic grid construction using Particle Swarm Optimization}
 
The grid-based approach previously described works in practice, but has an important computational cost and depends highly on the structure of the grid. Therefore, a key issue is the design of this grid of $\lambda$ values. Indeed, to estimate accurately the optimal $\lambda$ value, the grid must contain numerous elements. Even if the grid evaluations can be parallelized, the evaluation of one element does not consider the information obtained from previous evaluations to speed-up computation.  In this section, we propose a strategy to automatically build a sequence of $\lambda$ values that converges toward the value that minimizes BIC. As the gradient of the BIC with respect to $\lambda$ is intractable, the strategy must not be based on gradient calculation.  The proposed strategy is based on the principle of particle swarm optimization algorithms \citep{kennedy1995particle}, a zero-order optimization method. Particle swarm optimization is an iterative optimization algorithm in which candidate solutions, called particles,  are iteratively moved within the search-space toward the minimum of an objective function. At each iteration, the value of the objective function is evaluated for each particle's coordinate. Particles are then moved using simple rules in the direction of their own best-known position and the overall best-known position in the search-space. 
\\$ $\\
The objective function to minimize is the BIC. The particle swarm optimization algorithm considers $M$ candidate solutions whose coordinates in the search-space are different values of regularization parameters $\lambda = (\lambda_\beta,\lambda_\Gamma)\in \mathbbm{R}^{2}$. The initial position of each particle is initialized to the user-specified values $\{ \lambda_{0,1},\cdots, \lambda_{0,M} \}$. At each iteration $l<N_{PSO}$, particle's positions $\{ \lambda_{l-1,1},\cdots, \lambda_{l-1,M} \}$ are updated using three quantities:
\begin{itemize}
\item $\nu_{l-1,m}$, the actual speed of particle $m$
\item $\lambda_{P_m} - \lambda_{l-1,m}$, with $\lambda_{P_m}$ the position associated to the best values of particle $m$
\item $\lambda_{G} - \lambda_{l-1,m}$, with $\lambda_{G}$ the position associated to the overall best value
\end{itemize} 
The new position of each particle are then obtained using the following equations:
\begin{eqnarray}
\nu_{l,m} &=& \omega_{l}\nu_{l-1,m} + c_1(\lambda_{P_m} - \lambda_{l-1,m}) R_1 +c_2 (\lambda_G - \lambda_{l-1,m}) R_2 \nonumber \\
\nu_{l,m} &=& \mbox{Proj}_{  \left[ - \nu_{1}^{max} ; \nu_{1}^{max} \right] \times \left[ - \nu_{2}^{max} ; \nu_{2}^{max} \right] }( \nu_{l,m} ) \nonumber \\
\lambda_{l,m} &=& \lambda_{l-1,m} + \nu_{l,m} \nonumber
\end{eqnarray}
The constants $c_1$ and $c_2$ corresponds to the acceleration constants which control the speed the particles move toward the individual and the global best values. They are set to $c_1 = c_2 = 2$. This value is usually taken in most of the works using Particle Swarm Optimization and works well in practice \citep{marini2015particle}.  Quantities $R_1$ and $R_2$ corresponds to random vectors of size $2$ with each component uniformly distributed over $[0.1]$. The term $\omega_{l}$ is the inertia weight that control the influence of the velocity at the previous iteration on the current velocity. The value of $\omega_{l}$ is set according to a chaotic descending inertia weight \citep{marini2015particle}:
$$\omega_{l} =   \left( \bar{\omega}_{0} -  \bar{\omega}_{N_{PSO}} \right)\frac{N_{PSO} - l}{ N_{PSO} } + \bar{\omega}_{N_{PSO}}4U(1-U),\quad \mbox{ with } U\sim \mathcal(0,1), $$
parameters $\bar{\omega}_{0}$ and $\bar{\omega}_{N_{PSO}}$ are set to 0.9 and 0.4 respectively. The projection step is done to prevent "velocity explosion" \citep{marini2015particle} which may cause the divergence of the PSO algorithm. The maximal velocity value is set to $\nu^{max} = \frac{\lambda^{max}}{5}$ with $\lambda^{max} \in \mathbbm{R}^{2}$ the vector maximum authorized values for the regularization parameters. The algorithm is detailed in Algorithm \ref{algo:PSO} and schematically represented in the right part of Figure \ref{fig:SCHEMA_WR_PSO}.

At each iteration, the evaluation of each particle could be parallelized to speed up the algorithm. Moreover, warm restart can be used to decrease the number of iterations needed to achieve convergence for each particle evaluation. To evaluate $\lambda_{l,m}$ , the SAPG algorithm is initialized with $\hat{\theta}_{\lambda_{l-1,m}}$ and $S^{sa}_{\lambda_{l-1,m}}$ respectively the solution and the approximation of the conditional expectation of the complete data sufficient statistics obtained with the previous regularization parameters value $\lambda_{l-1,m}$. During the l-th iteration of the PSO algorithm, the sequence of stochastic approximation step sizes used during each SAPG run is the following:
$$\delta_{n}^{l}   =  \delta_\star \left(\frac{N_{SAPG}}{Freq} (l-1) + n \right)^{-\zeta} \mbox{ for } n \in \{1,...,N_{SAPG} \},$$
with $N_{SAPG}$, $Freq$ and $\delta_\star$ correspond respectively to the number of iterations of each SAPG run, the frequency of MCMC update and to the initial stochastic approximation step size used for the first PSO iteration.

\begin{algorithm}[H]
\caption{Particle Swarm Optimization Strategy}
\begin{spacing}{1.2}
\begin{algorithmic}[1]
\STATE \textbf{Initialization:} Set $\Lambda_0 = \{ \lambda^{1}_{0},\cdots, \lambda^{M}_{0} \}$, $\Theta_0 = \{ \theta^{1}_{0},\cdots, \theta^{M}_{0} \}$, $\mathbf{S}\textcolor{red}{^{sa}}_{0} = \{ S^{sa,1}_{0},\cdots, S^{sa,M}_{0} \}$, $\delta_0$, $\zeta$, $N_{SAPG}$
\FOR{$l \leq N_{PSO}$}
\FOR{$m \leq M$}
\STATE \textbf{1) Run SAPG}
\STATE $\quad$ $\theta_{l,m} = \mbox{SAPG}(\lambda_{l,m} \vert \theta_{l-1,m};S^{sa}_{l-1,m} )$
\STATE{\textbf{2) Update individual best}}
\IF{$BIC(\theta_{l,m})<BIC(\theta_{\lambda_{P_{m}}})$} \STATE {$\lambda_{P_{m}} = \lambda_{l-1}^{m}$} \ENDIF
\STATE{\textbf{3) Update global best}}
\IF{$BIC(\theta_{l,m})<BIC(\theta_{\lambda_{G}})$} \STATE {$\lambda_{G} = \lambda_{l-1}^{m}$} \ENDIF
\ENDFOR
\STATE{\textbf{4) Update $\mathbf{\lambda}$ values}}
\FOR{$m \leq M$}
\STATE{$R_1 \sim \mathcal{U}\left( [0;1]\right)^2$ and $R_2 \sim \mathcal{U}\left( [0;1]\right)^2$}
\STATE{$\nu_{l,m} = \omega_{l} \nu_{l-1,m} + c_1(\lambda_{P_{m}} - \lambda_{l-1,m}) R_1 +c_2 (\lambda_{G} - \lambda_{l-1,m}) R_2$}
\STATE{$\nu_{l,m} = \mbox{Proj}_{  \left[ - \nu_{1}^{max} ; \nu_{1}^{max} \right] \times \left[ - \nu_{2}^{max} ; \nu_{2}^{max} \right] }( \nu_{l,m} ) $}
\STATE{$\lambda_{l,m} = \lambda_{l-1,m} + \nu_{l,m}$}
\ENDFOR
\ENDFOR
\STATE{\textbf{Return $\lambda_{G}$ and $\theta_{\lambda_{G}}$ } }
\end{algorithmic}
\end{spacing}
\label{algo:PSO}
\end{algorithm}

The R code of the proposed method is available on the following repository: http://github.com/EdOllier/NLMEpen.

\section{Application to simulated data}\label{sec:sim}

\subsection{The two compartment pharmacokinetic model }\label{sec:model}

For the application to both simulated and real pharmacokinetics data, we used a two-compartmental pharmacokinetic model defined as:
\begin{eqnarray} \label{eq:A1C2}
  \frac{dA_{c}}{dt} &=&  D_{I}\mathbbm{1}_{t\leq T_{I}} + \frac{Q}{V_{p}}A_{p}  - \frac{Q}{V_{c}}A_{c}- \frac{Cl}{V_{c}}A_{c} , \\
  \frac{dA_{p}}{dt} &= & \frac{Q}{V_{c}}A_{c} -  \frac{Q}{V_{p}}A_{p} \nonumber,  
\end{eqnarray}
with $A_{c}(0) = D_{B}$, $A_{p}(0) = 0$. Variables $A_c$ and $A_p$ are respectively the amount of drug in the central and peripheral compartments. Quantities $D_{B}$, $D_{I}$ and $T_{I}$ are known and respectively corresponds to the bolus dose, the infusion rate and the duration of infusion. Here $V_c$ is the volume of the central compartment, $V_p$ is the volume of the peripheral compartment, $Q$ is the inter compartment clearance and $Cl$  is the elimination clearance. The plasmatic drug concentration of the i-th subject ($Y_{i,j}$, $1\leq j \leq J$) is then modeled as follows:
\begin{eqnarray}
&& Y_{i,j} = \frac{A_{c}(t_{i,j},Z_{i})}{{V_{c_i}}} +  \epsilon_{i,j}  \nonumber ,\\
&& \epsilon_{i,j} \sim \mathcal{N}(0,\sigma^2) \mbox{ (iid)}. \nonumber
\end{eqnarray}
The vector of individual parameters of subject i, $Z_i = ( V_{c_i}, V_{p_i}, Q_i, Cl_i) \in \mathbb{R}^4 $, is supposed to be a log-normally distributed 
$$ \log Z_i \sim \mathcal{N}\left(X_i \beta, \Omega \right).$$

Components of vector $\beta$ that corresponds to the intercept of the linear model of each parameter are noted $\mu = (\mu_{V_c},\mu_{V_p},\mu_{Q},\mu_{Cl})$ and are not penalized. This model was also used for the analysis of real data.

\subsection{Simulated data} \label{sec:simu}

Data were simulated using a two compartment pharmacokinetic model. The administrated bolus dose ($D$) was set to $1000$ $mg$ without infusion. The number of covariates was set to $K=50$ which corresponds to a model with a total of 215 parameters. Only two covariates were considered as influential. 

This corresponds to the following model for the individuals parameters : 
\begin{eqnarray}
 Z_i \sim \mathcal{N}\left( \left[ \begin{matrix}   1.82 + 0.4 \times x^{2}_{i} \\ 2.26 + 0.4 \times x^{2}_{i} \\ 3.10 \\ 1.67 + 0.4 \times x^{4}_{i}  \end{matrix}  \right], \Omega \right) 
 \quad \mbox{ with } \quad
 \Omega = \begin{pmatrix} 0.16 & 0 & 0 & 0.12 \\   0 & 0.3025 & 0 & 0 \\ 0 & 0 & 0.49 & 0 \\ 0.12 & 0 & 0 & 0.13 \end{pmatrix} .\nonumber
\end{eqnarray}

The vector of covariates for each subject was drawn from a centered Gaussian distribution with unit variance. To explore the impact of collinearity on the selection process, covariates were either simulated independently or based on a Toeplitz correlation structure with correlation parameter set to 0.8. Standard deviation of measurement errors was set to $\sigma = 5$. For each parameterization, we simulate $100$ datasets of $50$ and $100$ subjects with 7 observations per subjects at 6 min, 20 min, 45 min,1h, 2h, 4h and 8h after the bolus.

\subsection{Influence of adaptive step size on SAPG convergence}
To illustrate the benefit of the adaptive step size strategy on SAPG convergence, we applied the following step size sequence within SAPG :
\begin{itemize}
\item CSS: Common step size sequence (\ref{eq:gammaCst}). Different values for $\gamma_0$ were used: $\gamma_0 = 0.005$ (CSS-1), $\gamma_0 = 0.0025$ (CSS-2) and $\gamma_0 = 0.001$ (CSS-3).
\item MSS: Manually tuned step size sequence with a $\gamma_0$ specific to each type of component in $\theta$: $1.5 \times \gamma_0$ for $\beta$, $10 \times \gamma_0$ for $\sigma$ and $\gamma_0$ for $\Gamma$ and $\Delta$.  Different values for $\gamma_0$ were used: $\gamma_0 = 0.005$ (MSS-1), $\gamma_0 = 0.0025$ (MSS-2) and $\gamma_0 = 0.001$ (MSS-3).
\item ASS: Component-wise adaptive step size sequence (\ref{eq:gammaRandom}). Different values for $\gamma_0$ were used: $\gamma_0 = 0.05$ (ASS-1), $\gamma_0 = 0.025$ (ASS-2) and $\gamma_0 = 0.01$ (ASS-3)
\end{itemize}

\begin{figure}[H]
\begin{center}
\includegraphics[scale = 0.4]{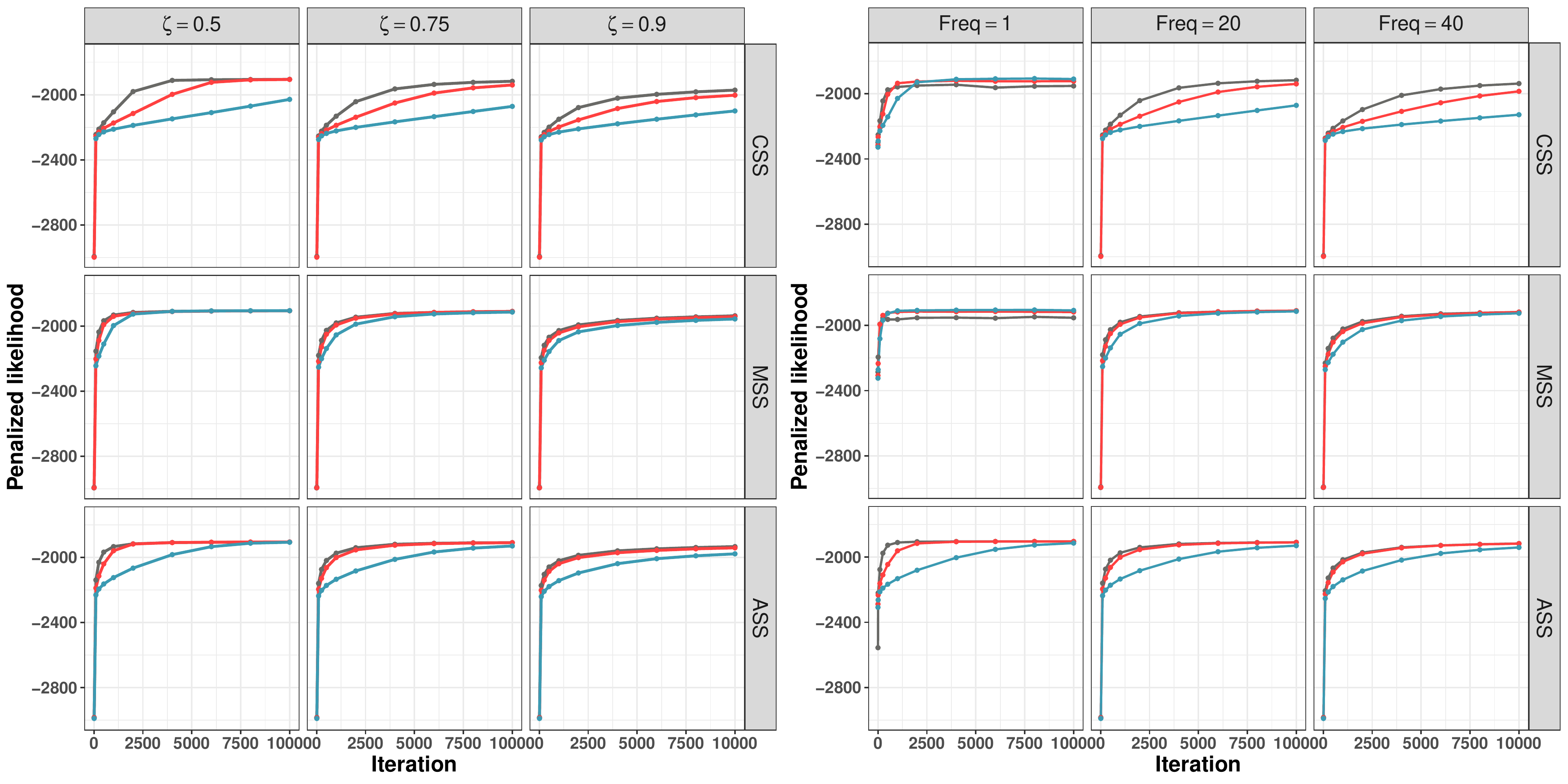}
\caption{Evolution of the penalized likelihood in function of iterations for the different step size tuning strategies (CSS, MSS and ASS). Grey lines correspond to CSS-1, MSS-1 and ASS-1. Red lines correspond to CSS-2, MSS-2 and ASS-2. Blue lines correspond to CSS-3, MSS-3 and ASS-3. Left panel corresponds to the evolution of the penalized likelihood for different values of $\zeta$ (0.9, 0.75 or 0.5) with the frequency of MCMC update set to 20. Right panel corresponds evolution of the penalized likelihood for different values of the frequency of MCMC update (1, 20 or 40) with $\zeta$ set to 0.75.}
\label{fig:SAPG_AS}
\end{center}
\end{figure} 

\begin{figure}[H]
\begin{center}
\includegraphics[scale = 0.5]{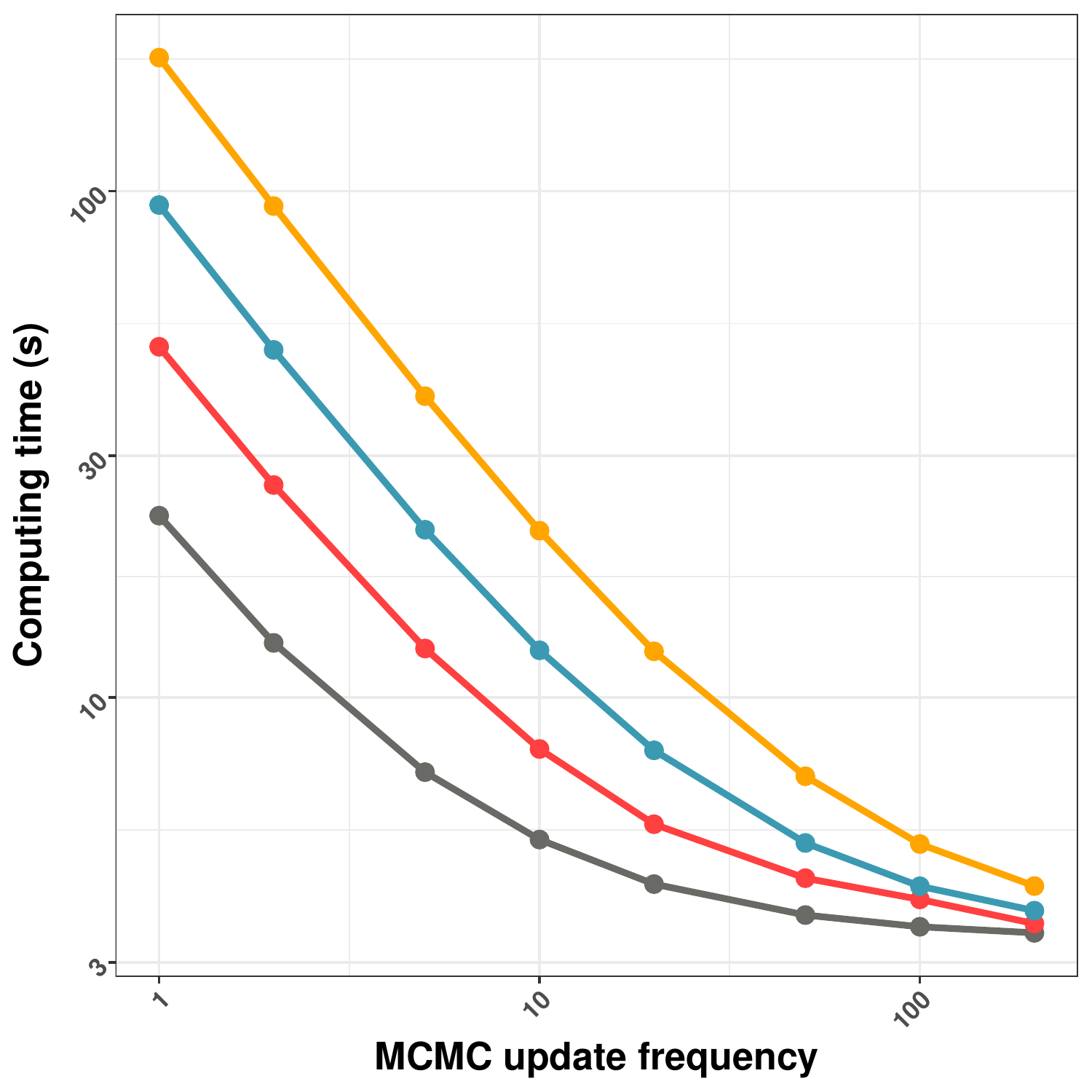}
\caption{Averaged time over 50 runs needed to perform $10^4$ iterations of SAPG in function of the MCMC update frequency. Colors correspond to different number of iterations used within each MCMC update. Black, red, blue and yellow correspond respectively to 2, 5, 10 and 20 MCMC iterations per MCMC update.}
\label{fig:SAPG_AS_TIME}
\end{center}
\end{figure}

For each type of step size sequence, SAPG was applied $50$ times to a simulated data based on the model described previously and the evolution of the penalized log-likelihood was evaluated:
$$ \ell(\theta_n)  - \lambda_\beta \Vert \beta_{n} \Vert_1 - \lambda_\Gamma \Vert \Gamma_{-_n} \Vert_{1}.  $$
 Figure \ref{fig:SAPG_AS} shows the evolution of the mean penalized log-likelihood in function of iterations obtained over 50 runs of SAPG for each step size tuning strategies (CSS, MSS or ASS) and each parameterization ($\gamma_0$, $\zeta$ and MCMC update frequency). Overall, manually tuned step size sequences (MSS) converged faster than sequences based on a common step size $\gamma_0$ for all the component's types. This illustrates the need of a component-wise adaptive step size sequence. Moreover, the CSS strategies are sensitives to the frequency of MCMC updates (right part of  figure \ref{fig:SAPG_AS}) with a decreasing convergence speed for a decreasing MCMC updates frequency. MSS strategy seems less sensitive to it. The ASS strategy convergence nearly as fast as MSS but without any hand tuning of $\gamma_0$ and is also robust to the changes in frequency of MCMC updates. By reducing the frequency of MCMC updates it is possible to reduce the computing time of SAPG as it is shown in figure \ref{fig:SAPG_AS_TIME}.
 
\subsection{Illustration of the PSO algorithm and the impact of warm restart}
To illustrate the advantage of the PSO algorithm over the grid search strategy and the impact of warm restart, we performed a simulation study. The following algorithms we applied to select the optimal model:
\begin{itemize}
\item PSO-WR: PSO algorithm with warm restart. Each SAPG run used a low number of iterations (4e3 iterations).
\item PSO-NoWR: PSO algorithm without warm restart. Each SAPG run used a low number of iterations (4e3 iterations).
\item Grid Search: user designed grid of 250 elements. Each SAPG run used a high number of iterations (12e4 iterations).
\end{itemize}
The grid search strategy was based on a uniform grid of 250 elements. For PSO-NoWR and PSO-WR, 25 particles were used. The number of iterations of the PSO algorithms ($N_{PSO}$) was set to 10 corresponding to a budget equivalent to the one used in the grid search strategy. For these three strategies, particles evaluations were performed in parallel using 3 CPU cores. 

\begin{figure}[H]
\begin{center}
\includegraphics[scale = 0.425]{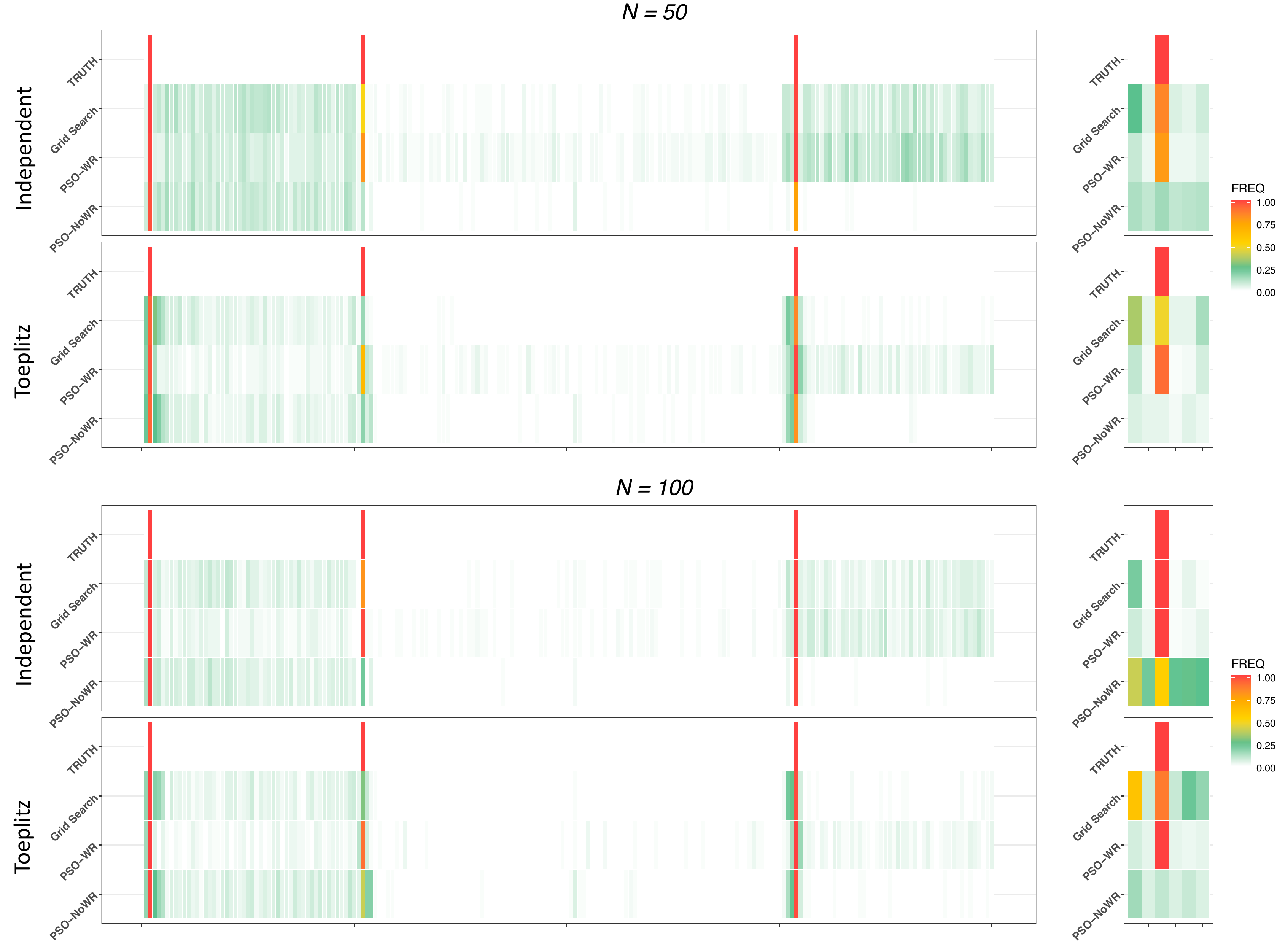}
\caption{Probability of being selected in the final model of each component of $\beta$ and $\Gamma$ with each algorithm over the 100 simulated datasets for both covariates design and the two sample sizes. Left: components of $\beta$. Right: components of $\Gamma$. }
\label{fig:PSO_SUPP}
\end{center}
\end{figure}

Figure \ref{fig:PSO_SUPP} illustrates the probability of being selected in the final model for each component of $\beta$ and $\Gamma$ with each algorithm over the 100 simulated datasets for both covariates design and the two sample sizes. Compared to the grid search strategy, PSO-WR better detected the support of $\beta$ and $\Gamma$. As shown in table \ref{table:PSO_SIM}, PSO-WR returned models with lower BIC values than the grid search strategy. This difference between the two approaches is even more important with the correlated covariates design. Using PSO-WR improved considerably the computation time of the selection procedure by reducing the number of iterations within each SAPG run. The median computing time was 59.3 min with the grid search strategy and 8.82 min with PSO-WR for datasets of 100 subjects. PSO-NoWR failed to correctly detect the support of $\beta$ and $\Gamma$ due to an insufficient number of iterations within each SAPG runs which preclude them to converge sufficiently. PSO-WR takes a bit longer to run than PSO-NoWR because of the additional iterations used in the SAPG runs of the first PSO iteration to ensure the convergence of quantities used to perform warm restart.

\begin{table}[H]
\footnotesize
\renewcommand{\arraystretch}{1.3}
\begin{center}
\begin{tabular}{| c  | c | c || c | c |  }
 \hline
  $ $ & \multicolumn{2}{c ||}{ BIC}    & \multicolumn{2}{c |}{ Time (min)}  \\
  \hline
 $ $ & \multicolumn{1}{c|}{N=50} &  \multicolumn{1}{c||}{N=100} & \multicolumn{1}{c|}{N=50} &  \multicolumn{1}{c|}{N=100} \\
  \hline
  $ $ & \multicolumn{4}{c |}{ Independent covariates}  \\
   \hline
  Grid Search & $2597 \mbox{ } [2568 ; 2617]$ &  $5154 \mbox{ } [5124 ; 5189]$  & $30.82 \mbox{ } [30.71 ; 31.33]$ &  $59.54 \mbox{ } [58.81 ; 59.79]$  \\
  PSO-NoWR & $2641 \mbox{ } [2618 ; 2661]$ &  $5237 \mbox{ } [5203 ; 5266]$  & $4.50 \mbox{ } [4.46 ; 4.60]$ &  $8.69 \mbox{ } [8.61 ; 8.79]$ \\
  PSO-WR & $2578 \mbox{ } [2558 ; 2601]$ &  $5142 \mbox{ } [5114 ; 5163]$  & $4.85 \mbox{ } [4.76; 4.91]$ &  $9.29 \mbox{ } [9.19 ; 9.39]$   \\
   \hline
     $ $ & \multicolumn{4}{c |}{ Correlated covariates}  \\
   \hline
  Grid Search & $2643 \mbox{ } [2621 ; 2668]$ &  $5250 \mbox{ } [5226 ; 5284]$  & $30.54 \mbox{ } [30.50 ; 30.66]$ &  $58.88 \mbox{ } [58.80 ; 59.51]$  \\
  PSO-NoWR & $2808 \mbox{ } [2645 ; 2610]$ &  $6239 \mbox{ } [3821 ; 3874]$  & $4.49 \mbox{ } [4.46 ; 4.61]$ &  $8.65 \mbox{ } [8.63 ; 8.90]$ \\
  PSO-WR & $2585 \mbox{ } [2565 ; 1905]$ &  $5144 \mbox{ } [5503 ; 6368]$  & $4.81 \mbox{ } [4.77; 4.93]$ &  $9.24 \mbox{ } [9.20 ; 9.48]$   \\
   \hline
 \end{tabular}
 \end{center}
\caption{Median and interquartile range of BIC values and computing times (min) obtained over the 2 covariates design and the 100 simulations for each selection procedure (Grid Search, PSO-NoWR and PSO-WR).}
\label{table:PSO_SIM}
\end{table}

\section{Application to real data : pharmacokinetics of tranexamic acid and cefazolin}
\label{sec:realdata}

We illustrate the proposed method on two real datasets describing respectively the pharmacokinetic of: i) tranexamic acid, an anti-fibrinolytic drug and ii) cefazolin, an antibiotic drug.The same two compartments model as in the simulation study was used for both dataset. Concerning tranexamic acid, the dataset contains data from 165 patients \citep{zufferey2017intravenous, lanoiselee2018tranexamic} with 5 samples per patient, sampled from $0$h to $8$h post administration. Each subject received a bolus of 1g at $t=0h$ and half of the subjects additionally received a perfusion of 1g during 8h. Concerning cefazolin, the dataset contains data from 73 patients (submitted work) with 4 to 5 samples per patient, sampled from $0$h to $8$h post administration. Each patient received a bolus of 2g at $t=0h$. In both datasets, 10 covariates were available: sex, age, body weight, glomerular filtration rate, ASA score, systolic blood pressure, diastolic blood pressure, hemoglobin, hematocrit, and history of congestive heart disease. All the covariates were standardized. Initially, the covariates and covariances were selected using a univariate testing method (one-step) for the tranexamic acid dataset and a stepwise method based on BIC for the cefazolin dataset.

The relevant covariates and correlations were selected by solving the penalized maximum likelihood problem (\ref{eq:PenLLAdapt}) using the SAPG algorithm. The optimal value for the sparsity parameters was calibrated by minimizing the $BIC$ using the PSO algorithm. The number of particles and the number of iterations were respectively set to 25 and 10.

\begin{figure}[H]
\begin{center}
\includegraphics[scale = 0.45]{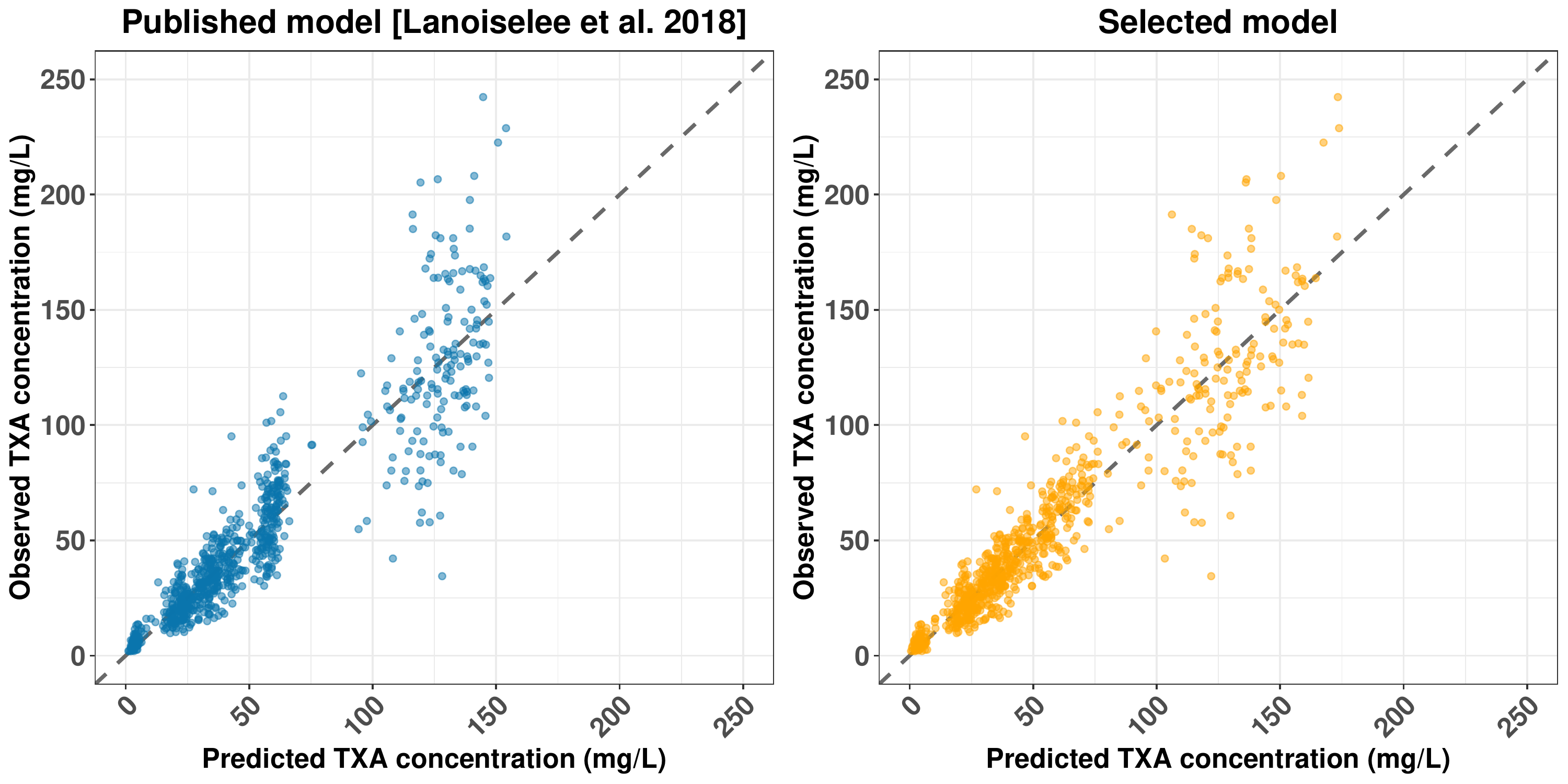}
\caption{Observed tranexamic acid concentrations versus predicted concentrations by both published and selected population models}
\label{fig:TXAObsPred}
\end{center}
\end{figure} 

The model selected for both drugs are compared in table \ref{table:MODEL_BIC} to the models published in the original publications. Parameter estimates of table \ref{table:MODEL_BIC} were obtained using  non standardized covariates values. For tranexamic acid, the model selected using the proposed approach contained 3 additional covariate parameters (BW on $V_P$ and $Q$, sex on $Q$) and no covariance contrary to the original published model \citep{lanoiselee2018tranexamic}. Inclusion of these 3 parameters resulted in a reduction of the BIC and the  unexplained between subject variability and in an improvement of the goodness of fit as shown in figure  \ref{fig:TXAObsPred}. These results have a direct practical interest, since the values of parameters $Q$ and $V_P$ are needed to compute an optimal bolus and maintenance dose designed to reach a desired target concentration. For example, the initial bolus dose to reach a target concentration $TC$ at t=0h could be calculated using the following equation based on the results of the initial publication:
$$ \mbox{Bolus dose} = TC\times \left( e^{\mu_{V_c} + \beta^{V_c}_{BW} \times (BW-80)} + e^{\mu_{V_p}}  \right).$$
The new formula based on the selected model considers the dependence of the parameter $V_p$ on body weight:
$$ \mbox{Bolus dose} = TC\times \left( e^{\mu_{V_c} + \beta^{V_c}_{BW} \times (BW-80)} + e^{\mu_{V_p} + \beta^{V_p}_{BW} \times (BW-80)}  \right).$$
For cefazolin, the selected model using the PSO algorithm was the same as the one originally selected (submitted work).
For comparison purpose, we also applied a stepwise method \citep{delattre2020iterative} on both datasets. It selected the same model as the penalized approach.

\begin{table}[H]
\footnotesize
\renewcommand{\arraystretch}{1.3}
\begin{center}
\begin{tabular}{ | c | c  c || c  c | }
  \hline
 $ $ & \multicolumn{2}{c||}{Tranexamic acid} &  \multicolumn{2}{c|}{Cefazolin} \\
 \hline
  $ $ & Published \citep{lanoiselee2018tranexamic} & Selected &  Published \citep{lanoiselee2021cefta} & selected \\
   \hline
 \multicolumn{5}{|c|}{ Fixed effects} \\
  \hline
  $\mu_{V_c}$ & $0.935$ & $1.089$ & $0.969$ & $0.969$  \\
   $\mu_{V_p}$ & $2.199$ & $1.277$ & $0.636$ & $0.636$  \\
    $\mu_{Q}$ &$3.318$ & $2.950$ &  $1.359$ & $1.359$  \\
     $\mu_{Cl}$ & $1.186$ & $1.185$ & $0.291$ & $0.291$  \\
           \hline
 \multicolumn{5}{|c|}{ Covariates} \\
  \hline
  $\beta^{V_c}_{BW}$ & $0.0111$ & $0.0097$ &  - & -  \\
  $\beta^{V_p}_{BW}$ & - & $0.0123$ &  - & - \\
  $\beta^{Q}_{BW}$ &- & 0.0061 & -  &  -  \\
    $\beta^{Q}_{Sex}$ &- & -0.330 & -  &  -  \\
  $\beta^{Cl}_{GFR}$ &$0.0056$ & $0.0055$ & $0.0084$  & $0.0084$ \\
      \hline
 \multicolumn{5}{|c|}{ Variances of random effects} \\
  \hline
  $\Omega_{V_c , V_c}$ & $0.130$ & $0.129$ & $0.353$  & $0.353$  \\
  $\Omega_{V_p , V_p}$ & $0.141$ & $0.087$ & $0.106$ & $0.106$  \\
  $\Omega_{Q , Q}$ &$0.120$ & $0.092$ & $0.534$ & $0.534$  \\
  $\Omega_{Cl , Cl}$ & $0.049$ & $0.051$ & $0.104$ & $0.104$  \\
        \hline
 \multicolumn{5}{|c|}{Covariances} \\
  \hline
  $\Omega_{V_c , Cl}$ & $0.026$ & - & $0.171$ & $0.171$ \\
          \hline
 \multicolumn{5}{|c|}{Residual error } \\
  \hline
  $\sigma$ & $4.01$& $4.01$ & $15.81$ & $15.81$ \\
    \hline
 \multicolumn{5}{|c|}{Bayesian Information criterion} \\
  \hline
    $BIC$ & $6607.6$& $6540.1$ & $3470.4$ & $3470.4$  \\
    \hline
 \end{tabular}
 \end{center}
\caption{Parameter estimates for the published and the selected pharmacokinetics models of tranexamic acid and cefazolin. For cefazolin, the published and the selected pharmacokinetic models were identical. Parameters estimates are provided using non standardized covariates values. Body weight was centered around 80 kg and glomerular filtration rate around 80 mL/min. Sex was equal to 0 for male and 1 for female.}
\label{table:MODEL_BIC}
\end{table}
 
\section{Discussion}
An automatic method to select covariates and correlation parameters in non-linear mixed effect model has been  proposed. In this approach, the space of candidate models is explored by solving a penalized likelihood problem for different regularization parameters values. Regularization parameters tune the level of sparsity within the model and are calibrated by minimizing the BIC. This optimization problem is solved using a particle swarm algorithm, a zero-order optimization method. It allows to solve the optimization problem without computing the gradient of the BIC with respect to the regularization parameters which is intractable. The use of warm restart enables to speed-up computation.

The problem of mixed effects models selection have been previously studied mostly for linear and generalized linear models \citep{bondell2010joint,schelldorfer2013glmmlasso}. Concerning nonlinear mixed effect models, Delattre et al. \citep{delattre2020iterative} have proposed a stepwise iterative algorithm based on the BIC to select both covariates parameters and random effects. Although it is a simple and effective method for low dimensional problems, this approach could be limited when the number of parameters increases as the dimension of the candidate models space grows exponentially. Although not evaluated in a high dimensional framework, the method proposed allows to handle a substantial number of parameters as illustrated in the simulations (215 parameters for 50 or 100 subjects). A limitation is indeed the absence of random effects selection. The selection of random effects using a penalized likelihood approach would necessitate to solve a non-convex and nonlinear optimization problem outside the exponential family. It is a much more difficult task from an algorithmic point of view. The optimization procedure would not rely anymore on the calculation of sufficient statistics but on the direct optimization of parameters within the structural model $f$ corresponding to a highly nonlinear function. However, the method could be applied to models with less random effects than parameters. For parameters without a random effect, a small and fixed (that is not optimized) random effect variance could be imposed by the user to allow the use of the proposed method.

Different extensions can be proposed. As in a more classical LASSO problem (\citep{tibshirani1996regression}), the use of adaptive weights may help to improve selection performance, especially in a high dimensional setting. To overcome this difficulty, the adaptive LASSO (\citep{zou2006adaptive}) has been proposed using a weighted $L_1$ norm penalty. In our case, a weighted estimator can then be defined, allowing the use of prior information:
\begin{equation*}\label{eq:PenLLAdapt}
\begin{split}
& \hat{\theta}_{\lambda_1,\lambda_2} =  \underset{\theta \in \mathbb{R}^d } {\operatorname{argmax}} \mbox{ }   \ell(\theta) - \lambda_\beta \Vert W_{\beta} \circ \beta \Vert_1 - \lambda_\Gamma \Vert W_{\Gamma} \circ \Gamma_{-} \Vert_{1}  \\ 
& \mbox{ subject to: } \sigma > 0, \quad \Delta \succ 0 
\end{split}
\end{equation*}
where $\circ$ is the Hadamard product, $W_\beta \in \mathbbm{R}^{(K+1)L} $ and $W_\Gamma \in  \mathbbm{R}^{L \times L}$ are penalty weights. For low dimensional problems, weights could be derived from unpenalized maximum likelihood estimates ($\hat{\beta}^{0}$, $\hat{\Gamma}^0$) following the adaptive lasso procedure (\citep{zou2006adaptive}): 
\begin{eqnarray*}
W_{\beta} =  \vert \hat{\beta}^0 \vert^{-\alpha} \mbox{ and } W^\Gamma_{i,j} =  \vert \hat{\Gamma}^0_{i,j}  \vert^{-\alpha}, \nonumber
\end{eqnarray*}
for some $\alpha>0$ (typically $\alpha=1$). For high dimensional problems, weights could be derived from a penalized maximum likelihood estimates ($\hat{\beta}^{\lambda_1}$, $\hat{\Gamma}^{\lambda_2}$): 
  \begin{eqnarray*}
  W_{\beta} =  \vert \hat{\beta}^{\lambda_1} \vert^{-\alpha} \mbox{ and } W^\Gamma_{i,j} =  \vert \hat{\Gamma}^{\lambda_2}_{i,j}  \vert^{-\alpha}, \nonumber
     \end{eqnarray*}
where parameters $\lambda_1$ and $\lambda_2$  are calibrated using the BIC. This adaptive strategy could be iterated and then corresponds to the multi-step adaptive lasso \citep{buhlmann2008}. An other extension is the use of structured penalties like group lasso or sparse group lasso \citep{simon2013sparse}.

To conclude, the proposed methodology provides an automatic approach to deal with model selection issues face in practice in domains such as pharmacokinetics. The fact that it is not restricted to very low dimensional covariates selection problem makes it interesting for emerging applications such as pharmacogenomics or pharmacoproteomics. 

\newpage

\bibliographystyle{plainnat}
\bibliography{BIB}

\end{document}